\documentclass[conference]{IEEEtran}
\usepackage{algorithmicx,algorithm}
\usepackage{algpseudocode}
\usepackage{cite}
\usepackage{amsmath,amssymb,amsfonts}
\usepackage{graphicx}
\usepackage{textcomp}
\usepackage{enumerate}
\usepackage{enumitem}
\usepackage[utf8]{inputenc}

\begin{document} 
\title{Edge-Assisted Congestion Control Mechanism for 5G Network Using Software-Defined Networking}
\author{\authorblockN{Meysam Nasimi\IEEEauthorrefmark{1}, Mohammad Asif Habibi\IEEEauthorrefmark{1}, Bin Han\IEEEauthorrefmark{1} and Hans D. Schotten\IEEEauthorrefmark{1}\IEEEauthorrefmark{2}
\IEEEauthorblockA{\IEEEauthorrefmark{1}Institute of Wireless Communication (WiCon), Technische Universit\"at Kaiserslautern\\ Paul-Ehrlich-Stra{\ss}e 11, 67663 Kaiserslautern, Germany\\Email: {\{nasimi, asif, binhan, schotten\}@eit.uni-kl.de}}}
	\IEEEauthorblockA{\IEEEauthorrefmark{2}Research Group Intelligent Networks, German Research Center for Artificial Intelligence (DFKI GmbH)\\Trippstadter Stra{\ss}e 122, 67663 Kaiserslautern, Germany\\
	Email: hans\_dieter.schotten@dfki.de}}
\maketitle
\begin{abstract}
In order to cope with the explosive growth of data traffic which is associated with a wide plethora of emerging applications and services that are expected to be used by both ordinary users and vertical industries, the congestion control mechanism is considered to be vital. In this paper, we proposed a congestion control mechanism that could function within the framework of Multi-Access Edge Computing (MEC). The proposed mechanism is aiming to make real-time decisions for selectively buffering traffic while taking network condition and Quality of Service (QoS) into consideration. In order to support a MEC-assisted scheme, the MEC server is expected to locally store delay-tolerant data traffics until the delay conditions expire. This enables the network to have better control over the radio resource provisioning of higher priority data. To achieve this, we introduced a dedicated function known as Congestion Control Engine (CCE), which can capture Radio Access Network (RAN) condition through Radio Network Information Service (RNIS) function, and use this knowledge to make the real-time decision for selectively offloading traffic so that it can perform more intelligently. Analytical evaluation results of our proposed mechanism confirm that it can alleviate network congestion more efficiently.
\end{abstract}
\section{Introduction} \label{sec:introd}
The vision of future Fifth Generation (5G) systems is to enable service delivery in ultra-dense networks. Particularly, always-connected devices, such as various types of smart phones, tablets, video-game consoles, Virtual/Augmented Reality (V/AR) devices and wearable electronics impose significant pressure on the backhaul and access networks. Moreover, the emerging Internet of Things (IoT) and massive Machine Type Communication (mMTC) are expected to introduce a huge number of machine connections~\cite{Cisco2014}. In this context, serious performances degradation in terms of QoS and/or Quality of Experience (QoE) is inevitable especially for the services with strict QoS requirements. Nevertheless, in such challenging environments, traffic bottlenecks in the core and backhaul networks can be reduced by locally processing data intensive task at the edge of the network in proximity to end-devices.

Mobile Cloud Computing (MCC) was introduced to deal with challenges of diverse and complex mobile service and application in terms of processing and data storage constraints in addition to battery lifetime, memory limitation and computational power of end-devices~\cite{Fernando2013}. MCC augmenting the resource capabilities of mobile devices by acting as an Infrastructure as a Service (IaaS) for data storage and processing. However, the MCC also imposes huge additional load both on radio and backhaul of mobile networks and introduces high latency since data is sent to powerful server that is far away from the users~\cite{Qi2012}.

To address the problem of long latency, the cloud services should be moved to a close proximity of the end users, i.e., to the edge of mobile network as considered in newly emerged edge computing paradigm. The edge computing can offer significantly lower latencies and jitter, mainly because the computing and storage resources are in proximity of the mobile users. Moreover, edge computing could exploit the contextual information for provisioning the network congestion states. This could indeed be achieved by combining MEC based application platform with the communication and context services that could be provided by potential 5G technologies \cite{Taleb2017}.

The European Telecommunications Standards Institute (ETSI) standard on MEC~\cite{Hu2015} may play an important role in this direction. MEC, as a key 5G network enabling technique, allows leveraging the cloud computing power by deploying application services at the edge of the mobile network. This can facilitate content dissemination within the access network. A key component for enabling MEC are servers integrated within the operator's RAN (e.g., 3GPP, Wi-Fi or small cells). MEC opens the door for authorized third parties, such as Content Providers (CP), to develop their own applications hosted in the MEC servers. These applications can add the flexibility to handle the traffic from/to mobile users. Besides, operators can expose their RAN edge Application Programming Interface (API) to authorized third parties to provide them with radio network information in a real-time manner.

The MEC framework consists of a hosting infrastructure and an application platform. The hosting infrastructure includes the MEC virtualization layer and the hardware components such as the computation, memory, and networking resources. The MEC application platform includes an IaaS controller together with the MEC virtualization manager, and provides multiple MEC application platform services. The MEC virtualization manager supports a hosting environment by providing IaaS facilities, while the IaaS controller provides a security and resource sandbox for both the applications and MEC platform. Four main categories of services are offered by MEC application platform including Traffic Offloading Function (TOF), RNIS, communication services and service registry.

In addition to MEC, Software-Defined Networking (SDN) paradigm, which is an emerging enabling technology, is utilized to facilitate data plane redirection mechanism through applying intelligence and centralize control over heterogeneous infrastructure~\cite{MadhusankaLiyanageAndreiGurtov2015}. Since the SDN controller has an overall view of the network, it has the visibility over data redirection. Furthermore, there are two important flow management protocols, known as OpenFlow \cite{Nunes2014} and Simple network Management Protocol (SNMP). Openflow is used for datapath control while SNMP is in charge of device control \cite{Yap2010}. In OpenFlow Wireless \cite{H1902,Araniti2014}, SDN can control network by adding protocol to BSs/APs software.

Additionally, Delay Tolerant (DT) traffic which accounts for a large portion of mobile data traffic is considered in this study. DT traffics are featured with relatively long latency in comparison with delay-sensitive traffics. For instance, e-mails, updates of social networking portals and firmware updates which can tolerate delay ranging from few seconds up to few minutes~\cite{Si2016}. However, DT traffic has its delay requirements or lifetime, which are much longer than delay sensitive traffic.

Significant research efforts have been invested on reducing the current overload of cellular networks. The most importantly, research works have analyzed the impact of traffic offloading and caching technique. Authors in~\cite{Sermpezis2015} propose an offloading mechanism in which the content can be delivered through small cells or Device-to-Device (D2D) communications. References~\cite{Mehmeti2013a,Mehmeti2014a,Mehmeti2017} investigate the performance of two type of WiFi offloading. The first one is on-the-spot offloading that is when there is WiFi available, all traffic is sent over the WiFi network; otherwise, all traffic is sent over the cellular interface. The second one is known as \textit{``delayed"} offloading where the traffic is delayed until WiFi connectivity becomes available. The work presented in~\cite{Duan2014} considers SDN-based WiFi data offloading, in which SDN controller facilitate the coordination between cellular and WiFi networks. In~\cite{Zeydan2016} and \cite{Bastug2014}, the role of proactive caching via small cells and D2D are investigated for 5G system. In particular,~\cite{Bastug2014} study the social networking and D2D use cases in order to exploit proactive caching. There has been some research in offloading computation to MEC or MCC~\cite{Guan2017,Liu2017,Messaoudi2017}. Considering the fact that MCC imposes huge additional load both on the RAN and backhaul and introduces high latency since data is sent to remote server. Therefore, MEC is seen as a promising approach to address the aforementioned problems. Moreover, MEC can provides an IT service environment and cloud-computing capabilities at the edge of the mobile network in close proximity to the users. While all the aforementioned studies present very attractive solutions, there are still limitations.

In this paper, we propose a congestion control mechanism in MEC context for reducing RAN congestion. The key idea is delaying of DT content from being delivered, until the delay conditions expire. This mechanism driven by the following two context factors:  i) the characteristic of data traffic (i.e., delay-tolerant data traffics) and ii) the network conditions (i.e., sudden traffic peaks).

The remainder of this paper is organized as follows: Section~\ref{sec:introd} introduces background and related work. Section~\ref{sec:model} presents the system model for the proposed mechanism. The congestion control mechanism is introduced in Section~\ref{sec:mechanism}, while Section~\ref{sec:perfanalysis} presents the analytical evaluation. The result and discussion is presented in section~\ref{sec:result-discussion}. Finally, the conclusion is drawn in Section~\ref{sec:conclusion}.

\section{System Model} \label{sec:model}
\begin{figure}
	\includegraphics[width=\linewidth]{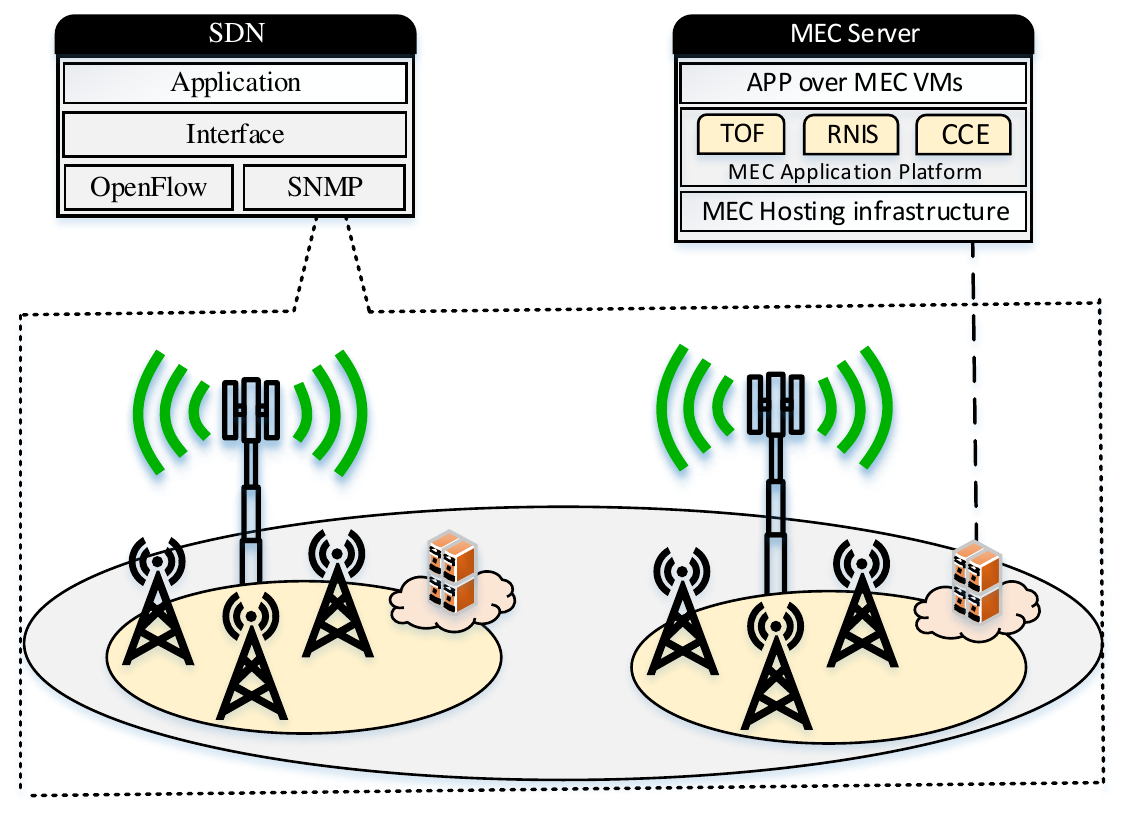}
	\caption{System Overview and module framework.}
	\label{fig:sysoverview}
\end{figure}

We consider a heterogeneous mobile network orchestrated by the SDN framework which is fully integrated with MEC~\cite{Chen2015}. This network is composed of Macro-cell Base Station (MBS), Small Cells (SC) and Mobile Node (MN). The MBS provides full coverage to subscribed MNs. The SCs are distributed within the MBS area to provide ample capacity to the few MNs within range. The system overview of this network is illustrated in Fig.~\ref{fig:sysoverview}.

In this work, the ETSI MEC~\cite{Hu2015} is considered as reference framework. It is assumed that there is a tight integration between SC and MEC in a way that a group of SCs are equipped with MEC server. Accordingly, MEC acts as an intermediate server so that DT contents can be temporarily stored and forwarded at later times. This significantly mitigates RAN load and improves resource utilization. Besides, the MEC server actively interacts with SDN through an API interface to facilitate traffic redirection. 

Additionally, we assume that the core network entities have some capabilities, which would enable them to classify traffics and then, based on the QoS requirements, assign a deadline (i.e., a maximum delay it can wait for) to each DT traffics~[12]. This can be achieved by leveraging Deep Packet Inspection (DPI) techniques. However, it should be noted that the traffic classification or DPI technique is not within the scope of this study.

Finally, It is assumed that MN traffics consist of two generic types of traffics namely, DT and delay sensitive traffics. Particularly, DT traffics refer to the type of traffics which are featured with long latency in comparison with that of delay sensitive. For instance, e-mails, updates of social networking portals and firmware updates~\cite{Si2016} can tolerate delay with range from few seconds up to few hours. Note that such DT traffic also has the delay constraints or lifetime. The only difference is that its tolerant delay is much higher than delay sensitive traffic.

\section{Congestion Control Mechanism} \label{sec:mechanism}
The goal of the proposed congestion control mechanism is to alleviate network congestion while makes better use of available network resources. A distinctive characteristic of our approach is that the MEC is playing an active role in this mechanism. The key idea is to intentionally delay DT content from being delivered and buffer it through an intermediate cloud server, with the goal of reducing RAN congestion, particularly during traffic peak hours.

We take advantage of RNIS cloud service introduced by ETSI, which is responsible for capturing real-time RAN condition. In addition, a dedicated function known as Congestion Control Engine (CCE) is proposed, which takes the RAN context information into account and perpetually monitoring the deadline of DT contents. 

With the proposed mechanism, a DT content is delivered depending on the network condition and their associated deadline. To illustrate, consider a situation that the network is overloaded, an operator can deliver DT content to an interested MN by temporarily storing content in MEC. In this context, the content will be transmitted to intended MEC over the backhaul and buffer it there until the congestion ratio is reduced under acceptable threshold or before the deadline expires.

The proposed algorithm consists of the following sequential steps (Algorithm~\ref{algo}):
\begin{enumerate}
\item\textbf{Packet inspection}: the network traffics are classified and then, each DT content assigned with a deadline based on their delay constraint.

\item\textbf{Congestion detection}: in order to identify RAN congestion, CCE constantly monitors RAN condition and in the case of congestion, it provide feedback to SDN.

\item\textbf{Redirection \& buffering}: for each successive time the network is found to be congested, SDN redirect the DT content to MEC storage where the content will be stored.

\item\textbf{Content delivery}: to capture the fact that buffered content may have a different deadline, CCE is monitoring the deadline of the content perpetually. If the deadline of a DT content is approaching, then the contents will abandon the storage and transmit to encountered requesters immediately. Otherwise, the content is kept until the RAN congestion is reduced to an acceptable level.
\end{enumerate}
Finally, note that mobile users prefer to have data immediately, However, they will be willing to accept delay for DT traffic (e.g. Email, software update, mobile backup, etc) if the mobile operator provides appropriate incentive in form of instantaneous price reductions~\cite{Sangtae2012}.

\begin{algorithm}[t]
\caption{Congestion Control Algorithm}\label{algo}
\begin{algorithmic}[1]
\State \textbf{Input} $deadline$ \Comment{Assigned Deadline}
\State \textbf{Input} $RA_{status}$ \Comment{RAN Congestion Condition}
\Procedure{CongestionControl}{$RA_{status}$,$deadline$}

\State $RA_{congested} =isRadioAccessCongested ()$
\While{($RA_{congested}$ = true)}
\State SDN redirect DT traffics $\rightarrow$ MEC.
\If{($deadline$ $\rightarrow$ expire)}
\State{deliver content immediately $\rightarrow$ requester.}\;
\Else
\State{keep content until deadline expires.}
\State{OR}
\State{keep content until congestion is relaxed.}
\EndIf
\EndWhile\label{euclidendwhile}
\EndProcedure
\end{algorithmic}
\end{algorithm}

\section{Analytical Evaluation} \label{sec:perfanalysis}
It is assumed that each MN is interested in different content over time. The content can deliver to an interested MN either by direct transmission from the MBS or transmitting the content to SCs over the backhaul. 

Two types of nodes are involved in this mechanism, requester of a content $R(t)$ and holder of content $H(t)$. The $R(t)$ is a MN that is interested in the content and not received it yet and the $H(t)$ is an MEC-assisted SC. The number of requesters, $r(t)$, shows how many users still need to be served at a given time. The number of holders, $h(t)$, represents the amount of resources used for serving user requests.

Concerning the holders of a content, we assume that MEC server stores the contents before their deadline expire, and during this time interval MECs always deliver them to encountered requesters through SCs. If a MN has been waiting for an amount of time, then the operator is obliged to deliver content before the expiration of their deadline. This is a reasonable assumption, since MECs are under the control of SDN, which knows the operating state of each MEC, and thus content discards (e.g., due to MEC overloads) can be avoided.

In order to analyze the performance of the proposed scheme, two key performance metrics are used. The first metric is the content delivery probability which represent how much traffic can be buffered in the edge server. The second metric is content delivery delay, which indicates that how fast content can be delivered~\cite{Sermpezis2017}.

The number of holders and requesters can be approximated over time through a mean filed approximation and a resulting system of ordinary differential equation. According to~\cite{Sermpezis2017}, the fluid-limit deterministic approximation for the expected number of holders $h(t)$ and requesters $r(t)$ at time $t$, is
\begin{equation}
h(t) = h(0) \cdot \frac{(r_0 + h_0)\cdot~e^{M_\lambda\cdot(r_0 + h_0)\cdot t}}{r_0 + h_0 \cdot~e^{M_\lambda\cdot(r_0 + h_0)\cdot t}}
\label{lem1}
\end{equation}

\begin{equation}
r(t) = r(0) \cdot \frac{r_0 + h_0}{r_0 + h_0 \cdot~e^{M_\lambda\cdot(r_0 + h_0)\cdot t}}
\label{lem2}
\end{equation}
where $h_0 = h(0^+)$ and $r_0 = r(0^+)$ at $t=0^+$, just after the initial placement of the content. $M_\lambda$ denotes the meeting rates (i.e., the edge nodes can exchange data only when they come within transmission range) between two nodes ${i,j}$ where $i \in MN$ and $j \in SCN$. Furthermore, the meeting rates $\lambda_{ij}$ are drawn from an (arbitrary) probability distribution $f_\lambda(\lambda)$ with mean value $M_\lambda$. Meeting duration is negligible compared to the time intervals between nodes, but long enough for a content exchange. 

\begin{figure}
	\includegraphics[width=\linewidth]{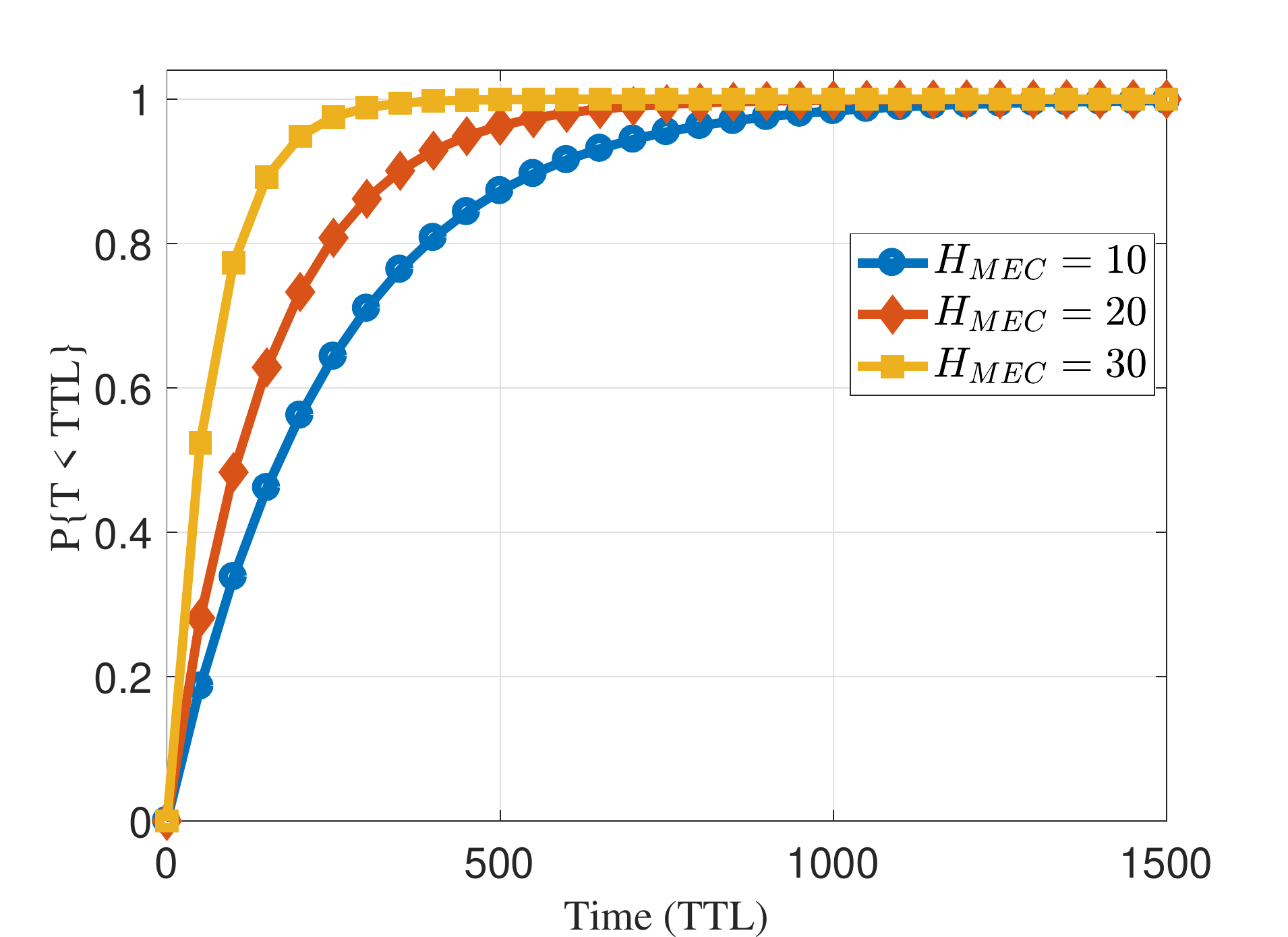}
	\caption{Delivery Probability $P\{T_d \leq TTL\}$ over time TTL. R(0)=50}
	\label{fig:del_probab}
\end{figure}

Based on~(\ref{lem1}) and~(\ref{lem2}), the desired performance can be calculated. Let us consider a requester $i \in r(0+)$, and denote as $T_i$ the time it receives the content. The probability that this (random) requester receives the content by a time $t$, i.e. $P\{T_i \leq t\}$, is equal to the percentage of offloaded contents by time t. Hence, we can write

\begin{equation}
P\{T_i \leq t\}=\frac{r_0 - r{(t)}}{r_0}= 1- \frac{r(t)}{r_0}
\label{eq1}
\end{equation}

Substituting the expression of~(\ref{lem1}) and~(\ref{lem2}) in~(\ref{eq1}), gives the following result for the content delivery probability.

The probability that a content is delivered by MEC to a user by time $t$ is given by

\begin{equation}
P_{dlv} = P\{T_d<t\}
\end{equation}

which can be written as fellow :

\begin{equation}
P_{dlv} = 1 - \frac{r_0 + h_0}{r_0 + h_0 \cdot e^{M_\lambda\cdot(r_0 + h_0)\cdot t}}
\label{new_equation}
\end{equation}

where $h_0$ and $r_0$ are number of content holders and requesters, respectively.

Finally, the expected content delivery delay, which represents the MN's experienced delay until it receives the content, is given by
\begin{equation}
E[T_d|TTL] = \frac{1}{M_\lambda\cdot~h_0} \cdot(1-\exp^{(-M_\lambda \cdot h_0 \cdot TTL)})
\end{equation}
where TTL denotes the assigned deadline.

\begin{figure}
	\includegraphics[width=\linewidth]{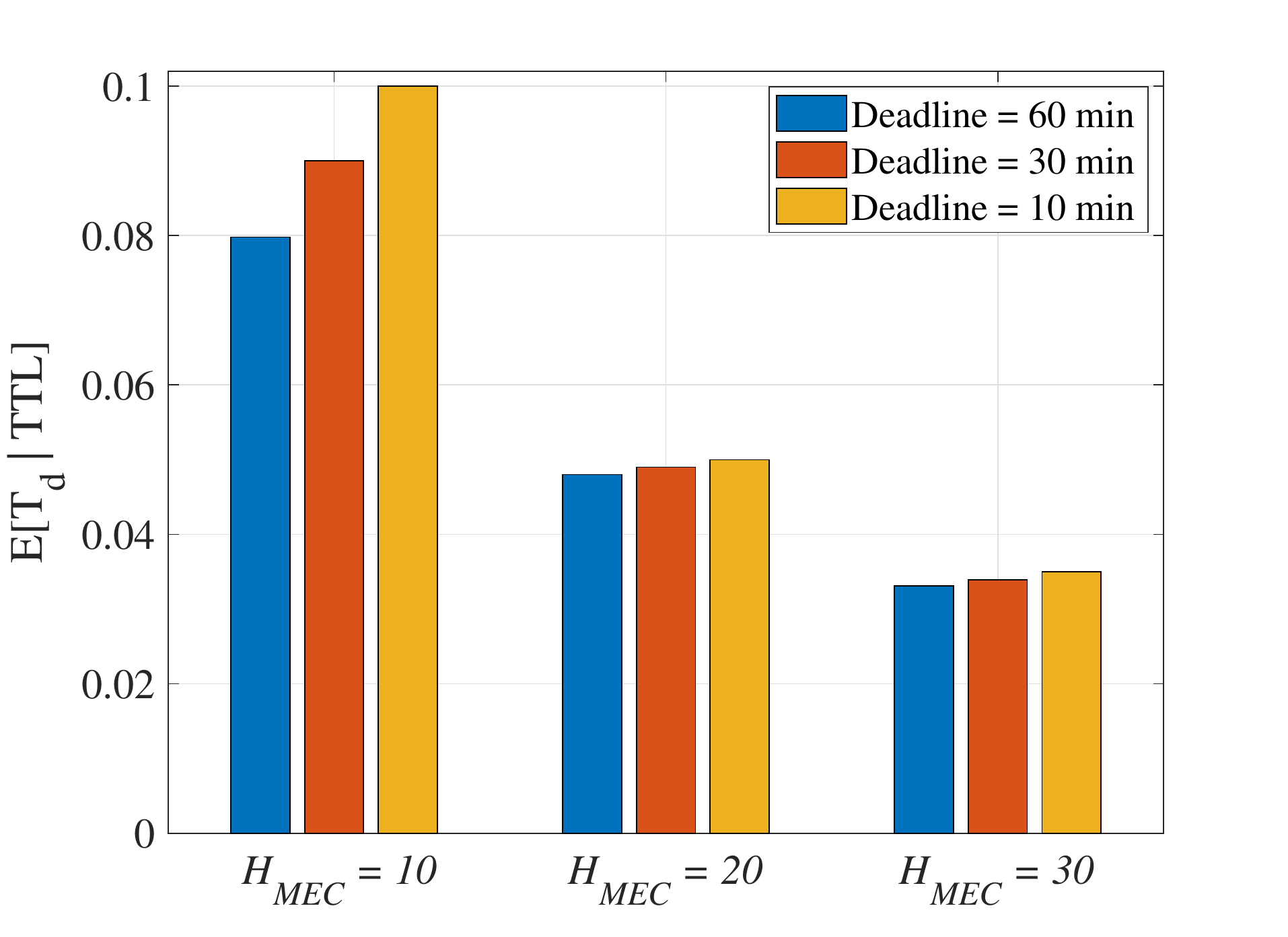}
	\caption{Expected delivery delay $E\{T_d | TTL\}$ for different deadlines.}
	\label{fig:delay_probab}
\end{figure}
\section{Results and Discussion} \label{sec:result-discussion}
In this section, we present analytical results to illustrate the performance of our proposed mechanism. In this respect, we consider two performance metrics, namely probability of content delivery and content delivery delay. We consider 100 MNs reside in the network with an average meeting rate $M_{\lambda}=3.3 * 10^-5$ meeting/sec~\cite{Hui2005}. The cellular network has to deliver DT contents to the MNs within deadline with range of 10 minutes, 30 minutes and 60 minutes.

Fig.~\ref{fig:del_probab} shows the delivery probability $P\{T_d \leq deadline\}$ of the content for the increasing number of the content holders (MEC server). Different density ratio of 10, 20 and 30 cloud servers are considered. It can be seen that increasing the number of content holders i.e. deploying more edge server, contribute to higher probability of content delivery for the short deadlines. This is more evident in a case that $H_{MEC}= 30$, where the RAN able to alleviate the content delivery within short deadlines.

Moreover, Fig.~\ref{fig:delay_probab} indicates the average delay a MN experiences until it receives the content in terms of the density of the edge servers for 10 servers, 20 servers and 30 servers. We compare the performance of the network with different deadlines of 10 minutes, 30 minutes and 60 minutes. It is clear from the figure that increasing the deadline of DT content yields lower content delivery delay. This is due to the fact that deploying more edge server can expedite content delivery by buffering more DT content especially during RAN congestion period, which result in lowering the content delivery delay.

\section{Conclusion} \label{sec:conclusion}
The explosion of data traffic has posed great challenges in terms of congestion and delay to the current networks. To cope with these two challenges, we have proposed an edge-assisted congestion control scheme which aims to alleviate network congestion in emerging 5G network environment. Supported by the MEC, the system is able to harvest context information for real-time RAN condition. Such knowledge is then translated into the dedicated function known as CCE to make decision for selectively buffering traffic. Performance evaluation results were presented to demonstrate the performance improvement of the proposed scheme.

\section{Acknowledgment}
This work has been performed with support of the H2020-MSCA-ITN-2015 project 5Gaura. The authors would like to acknowledge the contributions of their colleagues. This information reflects the consortia’s view, but the consortia are not liable for any use that may be made of any of the information  contained therein.

\bibliography{ISWCS2018.bib} 
\bibliographystyle{ieeetr}
\end{document}